\documentclass[a4paper,11pt]{article}
\usepackage{jheppub} 
\usepackage{lineno}
\usepackage{slashed}
\usepackage{amsmath}
\usepackage{xspace}
\usepackage{xcolor}
\RequirePackage{mathrsfs}

\newcommand{\unit}[1]{\,\mathrm{#1}}

\newcommand{\mev}{\unit{MeV}}
\newcommand{\gev}{\unit{GeV}}
\newcommand{\tev}{\unit{TeV}}

\newcommand{\figref}[1]{Fig.~\ref{#1}}
\newcommand{\secref}[1]{Section~\ref{#1}}
\newcommand{\appref}[1]{Appendix~\ref{#1}}
\newcommand{\eqnref}[1]{Eq.~\ref{#1}}

\renewcommand{\vec}[1]{\boldsymbol{#1}}
\renewcommand{\bar}{\overline}

\newcommand{\lvector}[2]{\begin{pmatrix}}
\newcommand{\micromegas}{\texttt{micrOMEGAs}}
\newcommand{\madgraph}{\textsc{MadGraph}\xspace}
\newcommand{\feynrules}{\textsc{FeynRules}\xspace}
\newcommand{\geant}{\textsc{Geant4}\xspace}
\newcommand{\egs}{\textsc{EGS5}\xspace}

\title{Production of Dark Photons through Higher Electromagnetic Moments at LDMX: Simulations and Model Discrimination}

\author[a]{Riccardo Catena}
\author[a]{Taylor R. Gray}
\author[a]{and Thomas Jerkvall}

\affiliation[a]{Chalmers University of Technology, Department of Physics, SE-412 96 G\"oteborg, Sweden}

\emailAdd{catena@chalmers.se}
\emailAdd{taylor.gray@chalmers.se}
\emailAdd{jerkvall@student.chalmers.se}

\abstract{We extend the projected sensitivity of LDMX for sub-GeV dark matter (DM) to the case of dark photons produced through higher order electromagnetic moments. These moments arise from loop diagrams involving portal matter fields, along with the gauge fields of new symmetry groups. Due to the Lorentz structures, in particular the momentum dependence, of these additional interactions, the kinematic distributions expected at missing momentum/energy experiments vary with model in addition to dark photon mass.
By considering four additional types of interactions -- magnetic and electric dipole, charge radius, and anapole moment -- we show that LDMX Phase-II is expected to probe the relic target of these additional dark photon models. We compare the analytic with the numerical methods for calculating the dark bremsstrahlung cross section, and compute the kinematic distributions for each model. The potential for model discrimination in the scenario of non-zero signal events at LDMX is discussed. We find that there is a degeneracy between the dark photon mass and model, which can be partially broken by considering both the energy and the transverse momentum of the recoil electron.
}

\begin{document}
\maketitle
\flushbottom
\raggedbottom

\section{Introduction}
\label{sec:intro}
The well established problem of the missing mass of the universe, based on undeniable evidence, has ignited the search for particles that can explain it. In more recent years, these Dark Matter (DM) searches have also been targeting masses outside of the canonical and widely explored Weakly Interacting Massive Particle (WIMP) \cite{Arcadi:2017kky,Roszkowski:2017nbc} mass window. The motivation to \textit{additionally} look beyond WIMPs is as follows \cite{Essig:2011nj}: The lack of discovery and ever shrinking available parameter space of WIMPs could be explained by it being too light, lighter than the mass of a nucleon, to induce a sizeable nuclear recoil at direct detection experiments. If we take that this lighter than WIMP DM particle is thermal -- was once in thermal equilibrium with the thermal bath -- a new mediator particle is required due to the Lee-Weinberg bound \cite{Lee:1977ua}. There are several proposed mediators that provide a channel between Standard Model (SM) fields and DM. Among them the dark photon, the gauge field of a hypothetical new $U(1)$ symmetry group, is the most widely studied \cite{Fabbrichesi:2020wbt,Cosme:2021baj, Catena:2023use, Curtin:2014cca,Graham:2021ggy}.
Hence, searches for this new mediator are complimentary to direct searches for probing these DM candidates. See \cite{Balan:2024cmq} for a review on sub-GeV DM with a dark photon mediator.

Fixed target experiments \cite{Antel:2023hkf,Ilten:2022lfq,Beacham:2019nyx}, such as the Light Dark Matter eXperiment (LDMX) \cite{LDMX:2018cma,Akesson:2022vza} and others \cite{NA64:2023wbi, Belle-II:2018jsg, COHERENT:2021pvd, SHiP:2020noy}, focus on $\mev - 
\gev$ mass DM, so-called sub-GeV DM. At these experiments, a relativistic electron or proton beam is incident on a fixed target giving rise to interactions between the target nuclei and the beam particles. Depending on the experiment, there are various ways a DM signal is measured such as in a downstream DM detector or by measuring the missing energy/momentum of the recoil electrons. For the purpose of this study, we consider the feasibility of the latter case corresponding to the upcoming LDMX. 
Final state electrons which have significant missing energy and have gained transverse momentum (above expected background) are considered signal events.
In this work, we are interested in signal events where 
a process analogous to ordinary photon bremsstrahlung occurs, dark bremsstrahlung \cite{Bjorken:2009mm}. 
Existing studies consider the production of dark photons \cite{Cline:2024qzv} produced by dark bremsstrahlung, which decay invisibly into DM of various different models \cite{Berlin:2018bsc,Catena:2023use,Rizzo:2021lob}.

The interaction between dark photons and SM fields emerge from vacuum polarization diagrams containing portal matter field loops \cite{Rizzo:2018vlb} -- leading to the ordinary kinetic mixing electromagnetic type interaction term.
As described in \cite{Rizzo:2021lob,Barducci:2021egn}, it is possible that the dark photon does not only couple through electromagnetic-like interactions, but also through higher order electromagnetic moment interactions. These additional interactions arise from loop diagrams with portal matter fields and new massive gauge bosons from a new gauge sector \cite{Rueter:2019wdf,Wojcik:2020wgm}.

Due to the different Lorentz structure of these interactions between the dark photon and the SM fermions, the expected number of events and kinematic distributions of these events vary from those of the electromagnetic type interaction. This opens the opportunity for potential model differentiation in the case of an LDMX signal, and an extension of the projected sensitivity to include these alternative models.
This concept is also employed in \cite{Blinov:2020epi}, where the authors consider the potential for model discrimination at missing momentum experiments.
In this study, we extend the work of \cite{Rizzo:2021lob} to investigate the signatures and projected sensitivity of dark photons with kinetic mixing, magnetic and electric dipole, charge radius, and anapole moment interactions with SM fermions. Taking DM to be a complex scalar field as a benchmark, which naturally evades strong limits from indirect detection and energy injection into the cosmic microwave background \cite{Cirelli:2023tnx,Slatyer:2015kla}, we calculate the thermal targets of these models in order to enforce the constraints from Planck \cite{Planck:2018vyg}, putting the sensitivities into context.
The analytic and numerical (using the Monte Carlo event generator \madgraph \cite{Alwall:2014hca}) approach to calculating the kinematic distributions are compared, where we find that the common analytic method over estimates the number of high transverse momentum events. 
We discuss the plausibility of model and dark photon mass differentiation in the optimistic case of enough signal events at LDMX. Introducing a new variable, the model, which the kinematic distributions are dependent on, invokes a more difficult mass estimation. However, even though we find a degeneracy between mass and model in the kinematic distributions, considering both energy and transverse momentum of the recoil electron breaks the degeneracy between certain models.

The theoretical framework for these additional dark photon - SM fermion interactions is presented in \secref{sec:theory}, which we follow by the discussion of the relic density calculations in  \secref{sec:relic}. The main results, namely the projected sensitivity and expected signal distributions are presented in \secref{sec:LDMX}. Finally in \secref{sec:conclusion} we leave our concluding remarks.

\section{Theoretical framework of Dark Moments}
\label{sec:theory}
The effective interaction vertex between spin-1/2 fields and a photon can be parameterized in terms of five electromagnetic form factors, consisting of all possible Lorentz invariant operators \cite{Nowakowski:2004cv}. These are effective interactions of higher order describing loop diagrams, inversely proportional to cut off scales of order of the loop field masses. The non-relativistic limit of these interactions reveal the static electromagnetic properties of the spin-1/2 field. The well studied kinetic mixing scenario of two $U(1)$ gauge groups \cite{Holdom:1985ag,Dienes:1996zr,DelAguila:1993px}, occurs through vacuum polarization diagrams containing new heavy portal matter fields carrying charges of both gauge groups.
After integrating out the portal matter fields, we arrive at the effective kinetic mixing Lagrangian commonly written down. As shown in Ref.~\cite{Rizzo:2018vlb,Rizzo:2021lob}, an analogous mechanism can generate 
new kinds of effective interactions beyond the common kinetic mixing scenario, including higher order electromagnetic moment interactions. Such higher order \textit{dark} electromagnetic moment interactions (also known as form factors) occurs similarly in the standard model \cite{Dombey:1980eu,Schwinger:1948iu,Muong-2:2003xkd,Hoogeveen:1990cb,Bernstein:1963qh}, and could occur in the dark sector between DM and the {\it ordinary} photon \cite{Barger:2010gv,Alves:2017uls,Lambiase:2021xcj,Pospelov:2000bq,Hambye:2021xvd}.

 We consider the scenario in which the SM is extended by a new $U(1)_D$ gauge group, with the associated gauge boson $A'$, the dark photon \cite{Fabbrichesi:2020wbt}, which receives a mass via the Stueckelberg mechanism or a dark Higgs mechanism. The existence of a new set of fields, portal matter fields\footnote{It has been shown in \cite{Rizzo:2018vlb} that the portal matter fields are likely fermions, and if so must be vector-like with respect to the SM and dark gauge groups.}, carrying SM and dark $U(1)_D$ charges allows for the interaction between the $A'$ and SM fermions through vacuum polarization diagrams with these portal matter fields in the loops \cite{Rizzo:2018vlb}.
 Previous literature has discussed examples of UV complete frameworks which contain these heavy portal matter fields \cite{Rueter:2019wdf,Rizzo:2018vlb,Wojcik:2020wgm,Rueter:2020qhf}.
As discussed in \cite{Rizzo:2021lob,Chu:2018qrm}, these loop diagrams generate additional interactions to the standard electromagnetic type interaction between $A'$ and SM fermions $f$. We consider in this study four additional types of interactions, separately\footnote{A scenario where $\epsilon$ is much smaller than the other effective interaction couplings is possible, since $\epsilon$ depends on the extended Higgs sector generating the mass of the dark photon \cite{Rizzo:2021lob}. Therefore we justifiably consider the implications for scenarios where each interaction is considered in isolation.}. We are justified to write these interactions in an EFT framework as higher dimensional interactions due to the small enough energy scale in which we are interested in. In general, the strengths of these interactions are dependent on the squared momentum of the $A'$, labeled $q^2$, and a cut off scale, $\Lambda_f$, where we can write the form factors as an expansion around $q^2=0$ as $c_0 + c_1 q^2/\Lambda_f^2 + \dots$, where $c_i$ are constants.
As described in \cite{Rizzo:2021lob} and calculated in \cite{Leveille:1977rc}, $\Lambda_f$ is dependent on the portal matter field masses. Since the masses of the portal matter must be at or above the TeV scale to evade LHC constraints, we set $\Lambda_f = 1 \tev$ for the purposes of this study.
We safely neglect the higher dimensional terms in the form factor, beyond $c_0$, since we are interested in sub-GeV masses, thus energies/momenta far enough below the TeV scale. We write these interactions as follows,
\begin{equation}
\begin{aligned}
    \mathscr{L}_{A'f} = & - e \epsilon Q_f \bar{f} \gamma^\mu f A'_\mu  
    - \mu_f \partial_\nu \left( \bar{f} \sigma^{\mu \nu} f \right) A'_\mu -i d_f \partial_\nu \left( \bar{f} \sigma^{\mu \nu} \gamma^5 f \right) A'_\mu \\ 
    &- b_f  \left[ \partial^\nu \partial_\nu \left(\bar{f}\gamma^\mu f \right) - \partial^\mu \partial_\nu \left(\bar{f}\gamma^\nu f \right) \right] A'_\mu \\ 
    &+ a_f \left[ \partial^\nu \partial_\nu \left(\bar{f}\gamma^\mu \gamma^5 f \right) - \partial^\mu \partial_\nu \left(\bar{f}\gamma^\nu \gamma^5 f \right) \right] A'_\mu,
\label{eq:lagrangian}
\end{aligned}
\end{equation}
where $f$ are the SM fermions\footnote{Coupling to neutrinos is assumed to be absent, and for the purposes of this study we mainly consider couplings to electrons and muons.} and $e$ is the elementary charge. The first term is the usual kinetic mixing term with interaction strength proportional to $\epsilon$, in addition to the magnetic dipole term with coupling $\mu_f$, electric dipole with $d_f$, charge radius with $b_f$, and anapole moment with $a_f$, which we take to be free parameters. We define the parameter $g_f$ representing the coupling strength for each type of interaction, or model,
\begin{equation}
g_f = 
\begin{cases}
    \epsilon & \text{Kinetic Mixing} \\
    \mu_f/e & \text{Magnetic Dipole} \\
    d_f/e & \text{Electric Dipole} \\
    b_f/e & \text{Charge Radius} \\
    a_f/e & \text{Anapole Moment} \\
\end{cases}
\end{equation}
where we compare the phenomenology of each. Notice that, for all cases except for kinetic mixing, $g_f$ is dimensionful with dimension -1 for the magnetic and electric dipole, and -2 for the charge radius and anapole moment.
The magnetic and electric dipole models are expected to have similar phenomenology, since their Lorentz structures only differ by a $\gamma^5$ thus their cross sections scale the same way with energy; similarly with the charge radius and anapole moment models. This allows for the models to be classified into three \textit{groups}: the first with the kinetic mixing, the second with the magnetic and electric dipole, and the third with the charge radius and anapole moment models.

In order to make meaningful statements on the phenomenology of these models, we take as a benchmark scenario, a complex scalar DM candidate, $\chi$, that constitutes the entire observed relic abundance and couples to the dark photon with the following Lagrangian,
\begin{equation}
    \mathscr{L}_{DM} = |\partial_\mu \chi|^2 - m_{\text{DM}}^2|\chi|^2 + ig_{\text{D}}A'^\mu \left[ \chi^* \left(\partial_\mu \chi \right) - \left(\partial_\mu \chi^* \right) \chi \right],
\end{equation}
where $m_{\text{DM}}$ is $O(\mev-\gev)$, $m_{A'}/m_{\text{DM}} = 3$, and we quantify in terms of the dark fine structure constant, $\alpha_D \equiv \frac{g_D^2}{4\pi}$. We take $\alpha_D = 0.5$, and $g_D \gg g_f$ such that $\text{BR}(A' \to \chi \chi) \approx 1$. 

Conservation of angular momentum forbids scalar particle s-wave annihilation via an s-channel dark photon --
giving rise to its p-wave dominant (velocity suppressed) thermally averaged annihilation cross section. This argument is independent from the type of interaction between the dark photon and SM fermions. Therefore, the scalar DM candidate is safe from strong indirect detection constraints in these models \cite{Boehm:2003hm, Cirelli:2021fbo}.

Direct detection of sub-GeV dark matter primarily relies on scattering with electrons or the excitation of vibrational modes in the detector material \cite{Essig:2024wtj}. As demonstrated in \cite{Rizzo:2021lob} for the magnetic moment model, the direct detection signal for electron recoils is expected to be small. We expect similar estimates for the electric dipole, charge radius, and anapole dark photon models. Although dedicated studies of the direct detection prospects for these higher electromagnetic moment dark photons present an intriguing direction for future study, we do not consider them in this work.

\section{Relic Density}
\label{sec:relic}

We enforce consistency with the Planck measurement of the DM relic abundance \cite{Planck:2018vyg}, 
\begin{equation}
    \Omega_{\text{DM,obs}} h^2 = 0.120 \pm 0.001,
\end{equation}
by solving the Boltzmann equation \cite{Gondolo:1990dk}, which calculates the relic abundance as a function of model parameters for each model,
\begin{equation}
    \dot{n}_\chi  + 3H n_\chi = -\frac{1}{2}\langle \sigma v_{\rm rel} \rangle_{\chi \chi \rightarrow f f}\, (n_\chi ^2 - n_{\chi ,{\rm eq}}^2) ,
\end{equation}
where $n_\chi$ is the cosmological number density of DM particles, $H$ is the Hubble rate, and $\langle \sigma v_{\rm rel} \rangle_{\chi \chi \to f f}$ is the thermally averaged cross section of DM annihilation into fermions, namely electrons and muons. $\langle \sigma v_{\rm rel} \rangle$ is calculated using the matrix elements presented in \appref{section:matrixelements}.

We use our own Boltzmann solver based on \cite{Gondolo:1990dk}, which is validated with \micromegas \cite{Belanger:2013oya} version 5. For the kinetic mixing model, we include contributions from DM annihilation into hadronic final states \cite{PDG,Ilten:2018crw,Izaguirre:2015zva}.
For the sake of this study, we consider large enough couplings, $\alpha_D = 0.5$, that DM reaches thermal equilibrium with the SM bath therefore the abundance is set through the freeze-out mechanism. Due to our choice of $m_{A'}/m_\text{DM} = 3$, DM annihilation into SM fermions through an s-channel with a virtual dark photon is the dominant process setting the abundance.
\begin{figure}
    \centering
\includegraphics[width=\linewidth]{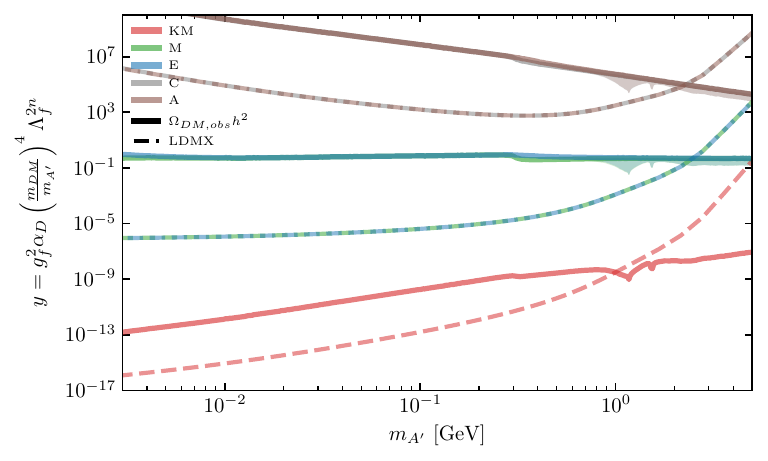}
    \caption{Projected sensitivity of LDMX \cite{LDMX:2018cma} of each model with their relic targets in the $y$ vs $m_{A'}$ plane, where $g_f$ is defined in \secref{sec:theory}. We take the set up of Phase II LDMX, which has $10^{16}$ EOT and an $8 \gev$ beam, for the optimistic zero background scenario at 90\% confidence level. We set $m_{A'}/m_\text{DM} = 3$, and $\alpha_D = 0.5$. The kinetic mixing (KM) is shown in red, magnetic dipole (M) in green, electric dipole (E) in blue, charge radius (C) in grey, and anapole moment (A) in brown, while the LDMX sensitivity is plotted with dashed curves and the relic targets with solid curves.}
    \label{fig:LDMX_sensitivity}
\end{figure}
\figref{fig:LDMX_sensitivity} plots the relic targets, the contours corresponding to the observed relic abundance in $y$ vs $m_{A'}$ space for each model in tandem with the projected exclusion bounds described in the following section. $y$ is defined to be,
\begin{equation}
    y \equiv g_f^2 \alpha_D \left( \frac{m_\text{DM}}{m_{A'}} \right)^4 \Lambda_f^{2n},
\end{equation}
where $\alpha_D$ and $g_f$ are defined in \secref{sec:theory}, and $n=1$ for the electric and magnetic dipole, and n=2 for the anapole and charge radius. We take an adapted version of the commonly used $y$ parameter, a convenient choice since the early universe DM annihilation cross section $\times$ velocity, $\sigma v_{\chi \chi \to f f} \propto \frac{y}{m_{\text{DM}}^2}$ \cite{LDMX:2018cma}. We include the factor $\Lambda_f^{2n}$ to have a dimensionless $y$ for all models.
For the higher order models we include a shaded region to the relic targets in \figref{fig:LDMX_sensitivity}, in which we expect contributions for hadronic final states. However, we do not explicitly compute these contributions for models beyond kinetic mixing interactions\footnote{The shaded regions in the relic targets are computed assuming Eq. A4 of \cite{Izaguirre:2015zva}. This computation is not valid for non-vector interactions, and we leave the development of a method to better approximate other types of dark photon interactions for future work. These shaded regions are only to be used as a proxy for the contribution from hadronic final states -- not to be taken rigorously.}.

As described in the previous section, the models can be separated into three groups, each leading to quantitatively similar relic targets: the first with the kinetic mixing, the second with the magnetic and electric dipole, and the third with the charge radius and anapole moment models. By examining the squared amplitudes of the process dominantly setting the relic density, found in \appref{section:matrixelements}, this will tell us about the thermally averaged cross section scaling with $m_{A'}$ near freeze-out. The relic target of the kinetic mixing model is validated with previous results \cite{Berlin:2018bsc}, with its positive slope. The magnetic and electric dipole models exhibit an approximately constant relic target as a function of dark photon mass. As discussed in \cite{Rizzo:2021lob} for the magnetic dipole model, which can also be applied to the electric dipole model, the energy-squared proportionality of the DM annihilation matrix element squared leads to a cross section near freeze-out that is approximately $m_{A'}$ independent. Finally, the charge radius and anapole moment model group's relic target slope is due to the additional energy squared proportionality in the matrix element squared, resulting in a cross section near freeze-out proportional to $m_{A'}^2$.

\section{Signatures at LDMX}
\label{sec:LDMX}

LDMX \cite{LDMX:2018cma}, a missing momentum experiment which uses an electron beam incident on a tungsten target, will have the ability to reconstruct both the energy, $E_{e^-_f}$, and the transverse momentum (momentum component perpendicular to the beam), $|p_T|_{e^-_f}$, of the recoil electrons. The events are expected to have different kinematic distributions whether or not there has been DM or dark mediator production, as opposed to the production of ordinary photons, since the production of an invisible but massive particle will carry away most of the energy and kick the recoil electron in the transverse direction. As we will show in this section, the nature of the DM signal also can vary the shape of the kinematic distributions and total expected number of events. Thus, the kinematic distributions will vary depending on the dark photon model, namely the type of interaction between the dark photon and SM fermions. 
The expected kinematic distributions are calculated analytically and numerically, with disagreement due to assumptions made in the analytic approach. We then proceed with the numerical approach, and compute the projected sensitivity in reference to our calculated thermal targets.

\subsection{Analytic vs Numerical Methods}

Using the formalism presented in \cite{Bjorken:2009mm}, which relies on the Weizsacker-Williams Approximation \cite{vonWeizsacker:1934nji,Williams:1935dka}, and comparing with the results of \cite{Rizzo:2021lob}, we calculate the $E_{e^-_f}$ and $|p_T|_{e^-_f}$ distributions, namely $\frac{d \sigma}{dx_{A'}d\cos\theta}$, in the kinetic mixing and magnetic dipole models.
The differential cross section for dark photon production in the kinetic mixing model is given by \cite{Bjorken:2009mm},
\begin{equation}
    \frac{d \sigma_{KM}}{dx_{A'}d\cos \theta} = A x_{A'} \left[ \frac{1-x_{A'}+x_{A'}^2/2}{U^2} + \frac{(1-x_{A'})^2 m_{A'}^2}{U^4}\left( m_{A'}^2 - \frac{U x_{A'}}{1-x_{A'}} \right) \right],
\end{equation}
where $A$ is a constant which depends on $\epsilon$, $E_{\text{beam}}$, an effective flux of photons depending on the nuclear structure, and $m_{A'}$, $x_{A'} = E_{A'}/E_{\text{beam}}$, $\theta$ is the angle of the dark photon with the beam axis, and $U = E_\text{beam}^2 \theta^2 x + m_{A'}^2 \frac{(1-x)}{x}+m_e^2 x$.
Similarly, from \cite{Rizzo:2021lob} we have the differential cross section for dark photon production in the magnetic dipole model,
\begin{equation}
    \frac{d \sigma_{M}}{dx_{A'}d\cos \theta} = \frac{B x_{A'}^2}{U},
\end{equation}
where $B$ is a constant. Notice that the differential cross sections do not depend on parameters associated with the DM particle -- only the dark photon mass and coupling to electrons. The distributions of the recoil electron are computed by integrating $ \frac{d \sigma}{dx_{A'}d\cos \theta}$ over $\theta$ from 0 to $\pi$, and normalizing to properly compare with the numerical method described below.
These analytic expressions rely on a number of assumptions. The dark photon is assumed to be roughly collinear with the beam, hence $\theta$ is small. In addition, the nucleus is assumed to remain at rest i.e. not recoil at all during this process.

Keeping these assumptions of the analytic method in mind, we simulate this dark bremsstrahlung process by running \madgraph \cite{Alwall:2014hca} with our \feynrules \cite{Alloul:2013bka} generated UFO files for each model. We include the form factors for the nucleus-photon vertex given in A18 and A19 of \cite{Bjorken:2009mm}\footnote{As pointed out by , there is a typo in eq. A19, where the second term should not be squared.}, which include both elastic and inelastic contributions. The signal contribution from a dark photon emitted off of the nucleus rather than the electron is safely neglected, since we expect the overall number of events and generated $p_T$ to both be small, however we leave a detailed study of the contribution to future work.
\begin{figure}[h!]
    \centering
    \includegraphics[width=0.8\linewidth]{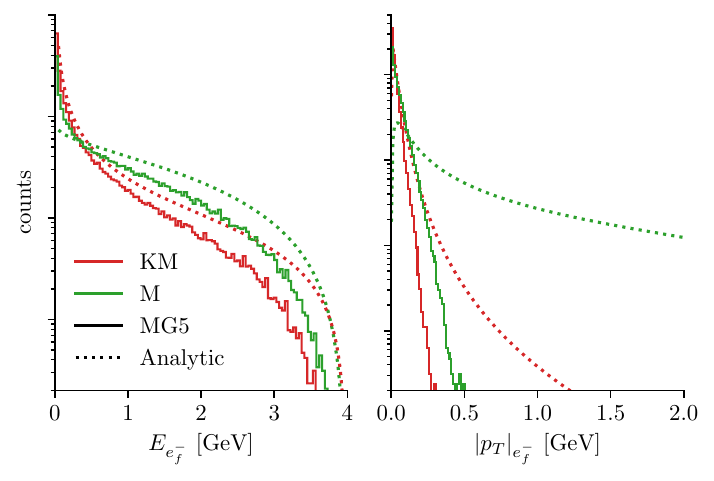}
    \caption{Kinematic distributions, calculated numerically with \madgraph (solid) and analytically (dotted), of the recoil electron, $e^-_f$, in both the kinetic mixing (KM, red) and magnetic dipole (M, green) models. The process simulated and calculated is dark photon bremsstrahlung with $m_{A'}=100 \mev$ and $E_{\text{beam}} = 4 \gev$. Each histogram is normalized, thus the relative bin heights should not be analyzed, rather the difference in shapes.}
    \label{fig:kinematics_analytic}
\end{figure}
The comparison between the analytic and numerical methods is plotted in  \figref{fig:kinematics_analytic}, for both the $p_T$ and $E$ distributions of the recoil electron. For the $E$ distributions, the shapes of the analytic distributions for both models resemble the numerical distributions. In fact, altering the upper integration bound over $\theta$ to a value closer to zero brings the analytic curves closer to the numerical curves. However, the discrepancy is evident for the $p_T$ distribution, where for the analytic method there are more events with higher $p_T$. This is due to faults in the assumptions -- in particular events with higher $p_T$ are overestimated since the nucleus is assumed to always be at rest, unable to take any of the transverse momenta generated from the dark photon.
\begin{figure}[h!]
    \centering
    \includegraphics[width=0.6\linewidth]{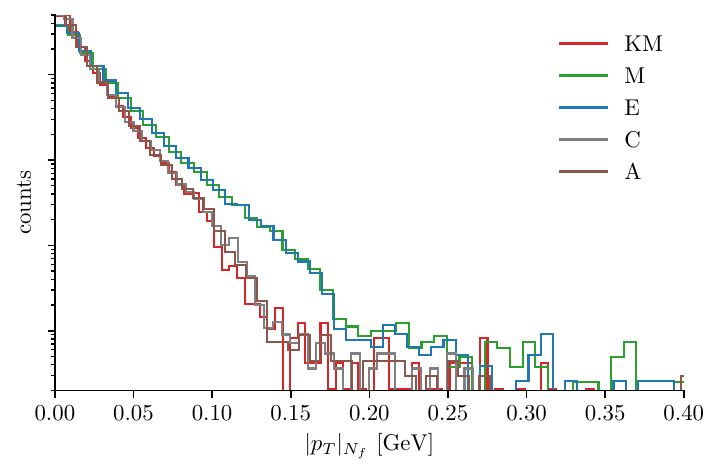}
    \caption{Distribution of final state tungsten nucleus transverse momenta $p_T$, for each model, simulated using \madgraph.}
    \label{fig:nucleus_pT}
\end{figure}
The $p_T$ distribution of the final state tungsten nucleus, directly after the process, is plotted in \figref{fig:nucleus_pT} using \madgraph. There are a significant number of events where the nucleus receives $>100\mev$ transverse momenta, a non-negligible quantity not accounted for in the analytic method.
The momentum transfer to the nucleus is also neglected in the software packages \geant \cite{GEANT4:2002zbu} and \egs \cite{Hirayama:2005zm} for modeling bremsstrahlung in the SM. In \cite{Blinov:2024pza}, the authors similarly point out problems with this assumption.
For this reason, we continue this section using \madgraph to simulate dark bremsstrahlung events for each model, computing the projected sensitivity and expected kinematic distributions potentially used for model discrimination in case of a future signal at LDMX.

\subsection{Projected Sensitivity}
The projected exclusion bounds at 90\% confidence level of LDMX are plotted in \figref{fig:LDMX_sensitivity} for each model. We compute the projected exclusion bounds using \madgraph simulations of dark bremsstrahlung events at LDMX phase II which plans to have $10^{16}$ electrons on target (EOT) and a beam energy of $8 \gev$. Assuming that the events follow Poissonian statistics and the optimistic zero background case, we draw our projected exclusion bound at 2.3 events. The relic targets are included, calculated as described in \secref{sec:relic}, showing the ability to probe each model in a region of parameter space compatible with relic abundance measurements. 
 Similarly to the relic targets, the projected exclusion bounds vary for each model due to the different Lorentz structures in the $A'-f$ vertices and can be classified into three groups. Each group has a different expected number of events, and $m_{A'}$ proportionality of the dark bremsstrahlung cross section.
 
\subsection{Kinematic Distributions}
 As mentioned above, and shown in \cite{Rizzo:2021lob}, different models lead to different distributions of the recoil electron at missing momentum experiments, such as LDMX.
\begin{figure}[h!]
    \centering
    \includegraphics[width=\linewidth]{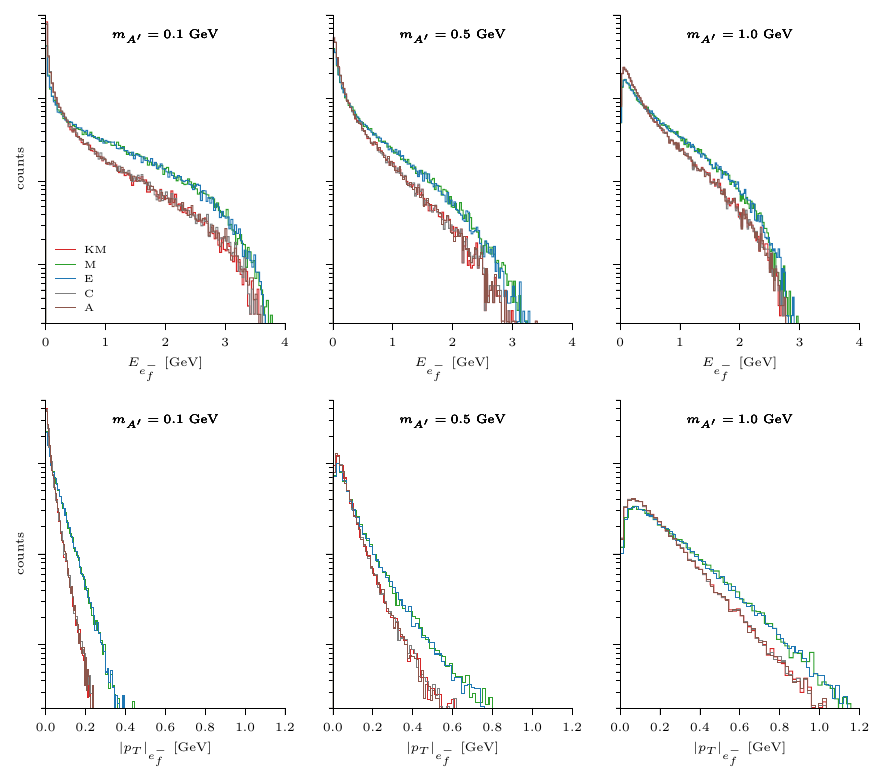}
    \caption{Kinematic distributions, the energy, $E_{e^-_f}$, and the transverse momentum $|p_T|_{e^-_f}$, of the recoil electron from MG simulations of a $4 \gev$ $e^-$ beam LDMX set up. The kinetic mixing model is drawn in red, magnetic dipole in green, electric dipole in blue, charge moment in grey, and anapole moment in brown.}
    \label{fig:LDMX_distributions}
\end{figure}
For three representative dark photon masses, the transverse momentum and the energy of the final state electron, $e^-_f$, are plotted in   \figref{fig:LDMX_distributions} for each model. Similarly to the projected sensitivity curves, \figref{fig:LDMX_sensitivity}, the magnetic and electric dipole models are qualitatively similar to eachother, as well as the charge radius and anapole moment models. However, the kinetic mixing model distributions have a shape similar to the distributions of charge radius and anapole moment: from \eqnref{eq:lagrangian}, since $q^2 = m_{A'}^2$ in the on-shell dark photon bremsstrahlung, there is no longer momentum dependence in the Feynman rule for the charge radius and anapole models, just as for kinetic mixing. The overall scaling of the kinetic mixing model events would be much larger, however the distributions have been normalized - bringing the number of groups from three to two.
The difference in shapes between the two groups arises from the momentum dependence of the Feynman rules.
 The charge and anapole models have momentum dependent vertices, where the Feynman rule is dependent on $q$, the momentum of the dark photon.
 
 \begin{figure}[h!]
    \centering
    \includegraphics[width=\linewidth]{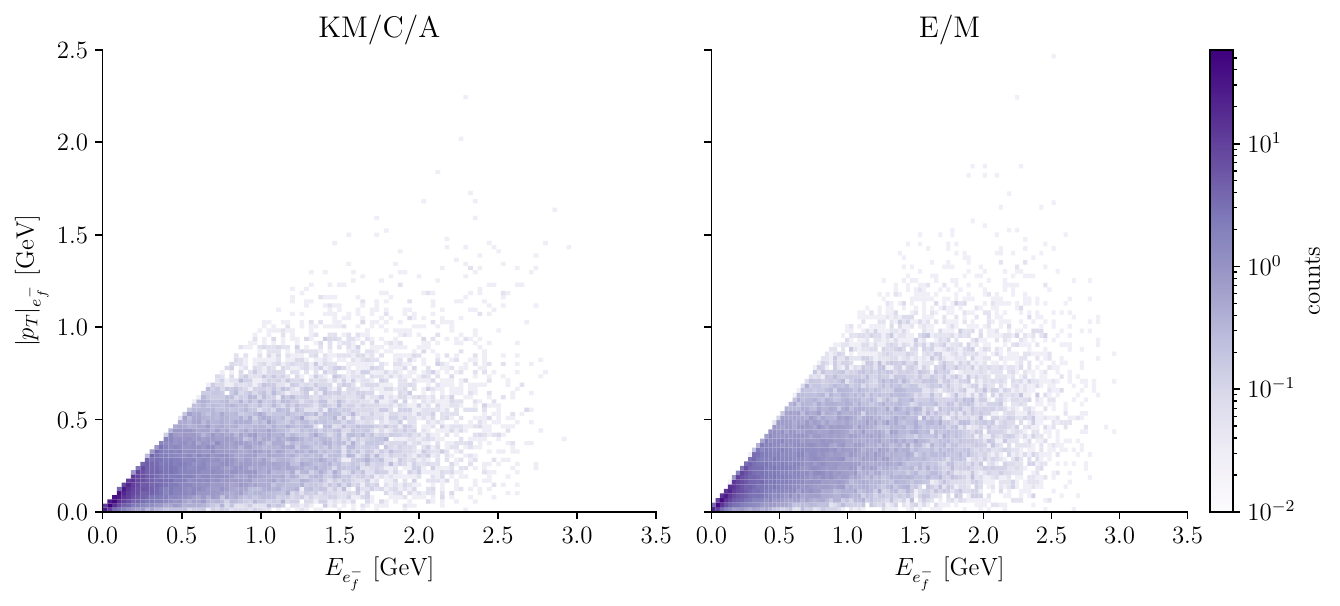}
    \caption{The normalized kinematic distributions of \figref{fig:LDMX_distributions}, plotted as 2-dimensional histograms in the $|p_T|_{e^-_f}$ vs $E_{e^-_f}$ plane. The kinetic mixing, charge radius, and anapole moment models are plotted on the left, while the magnetic and electric dipole models are on the right; where the two groupings are chosen since the distributions are the same within them apart from statistical fluctuations. The dark photon mass is set to $m_{A'} = 1 \gev$.}
    \label{fig:LDMX_2D}
\end{figure}
\figref{fig:LDMX_2D} plots the distributions of \figref{fig:LDMX_distributions}, but as a 2-dimensional histogram, highlighting the correlations between the two kinematic parameters measurable at LDMX. As evident by \figref{fig:LDMX_distributions} and already discussed above, the models can be grouped by their signal behaviour in which their distributions are the same apart from statistical fluctuations. Therefore, since the distributions for the models belonging to each group are the same, neglecting statistical fluctuations, the distributions are presented as one plot per group. The left and right 2-dimensional distributions indeed do differ; there are more events towards higher energy and $p_T$ for the magnetic and electric dipole models. All events obey $E_{e_f^-} > |p_T|_{e_f^-}$, as expected, and most have energy and $p_T$ $\sim 1 \gev$.

\subsection{Degeneracy of Mass and Model}

Considering the ordinary kinetic mixing dark photon model with enough signal events, the dark photon mass can be estimated based on the shape of the kinematic distributions \cite{LDMX:2018cma,Akesson:2022vza}. This is possible since the only model parameter altering the shape of the distributions is $m_{A'}$. Interestingly, considering higher order electromagnetic moment interactions in addition to the standard kinetic mixing, complicates the mass estimation process, since varying the model also alters the kinematic distributions.
 Increasing $m_{A'}$ pushes the $E_{e^-_f}$ distribution closer to zero and the $|p_T|_{e^-_f}$ to larger momenta, and on the other hand the magnetic and electric dipole moment models give rise to more higher energy and $p_T$ recoil electrons.
Therefore, varying the model can have indistinguishable effects from varying the mass.
\begin{figure}[h]
    \centering
    \includegraphics[width=\linewidth]{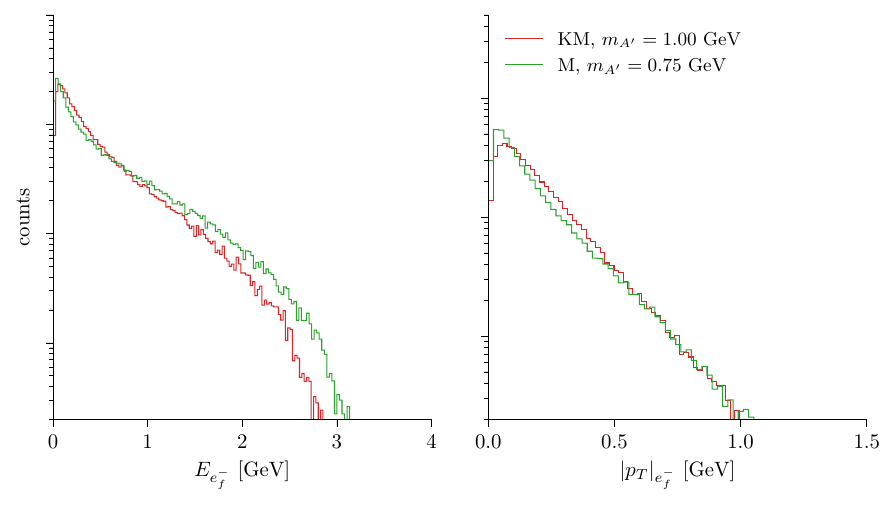}
    \caption{Kinematic distributions of the recoil electron for the kinetic mixing model at $m_{A'}=1.0 \gev$ (red) and the magnetic dipole model at $m_{A'}=0.75 \gev$ (green).}
    \label{fig:kinematic_massgen}
\end{figure}
\figref{fig:kinematic_massgen} shows an example of a case where two different dark photon masses can result in similar kinematic distributions of the recoil electron if the models are different. Both of the $|p_T|_{e^-_f}$ distributions, plotted in the right subfigure, exhibits a degeneracy in the mass and model. Therefore, one could falsely fit the mass to be 1 GeV when in fact the mass is actually 0.7 GeV but with a magnetic dipole type interactions rather than standard kinetic mixing. 

Since the $|p_T|_{e^-_f}$ distribution varies with respect to $m_{A'}$ in a different way compared to the $E_{e^-_f}$ distribution, the degeneracy can be broken to an extent. As evident by the $E_{e^-_f}$ distribution on the left, the two distributions have a visible separation. Therefore, including both the $|p_T|_{e^-_f}$ and $E_{e^-_f}$ distributions can break some of the degeneracies, at least between the two groups, where the first group consists of kinetic mixing, charge radius, and anapole moment models, and the second of magnetic and electric dipole models. 
Distinguishing between models within each group would require additional experiments and considerations if at all possible, which is beyond the scope of this work. The only exception to this is that one can distinguish between the kinetic mixing model and charge radius or anapole moment models, since the overall expected number of signal events is higher for kinetic mixing -- as evident by \figref{fig:LDMX_sensitivity} which shows the varying magnitudes of the sensitivity, thus number of signal events, between the models.

\section{Conclusion}
\label{sec:conclusion}

We consider an extended theoretical framework for the dark photon, where there are effective interactions involving loops of portal matter fields, in addition to the well studied electromagnetic-like kinetic mixing interaction, as presented in \cite{Rizzo:2021lob}.
Due to the different Lorentz structures of the $A'-f$ interaction vertices leading to different energy dependencies in the dark bremsstrahlung and DM annihilation cross sections, the expected signal distributions of LDMX and relic targets vary between models. 
By studying each new type of interaction separately, we compute the kinematic distributions expected at LDMX, the projected sensitivities of LDMX, and the relic targets -- extending the theory reach of LDMX to include these additional models. Implications for other future and current experiments is reserved for future studies. We find that the models can be classified into three groups, each characterized by similar vertices thus quantitatively similar expected LDMX signals and relic targets: the first with the kinetic mixing model, the second with the magnetic and electric dipole models, and the third with the charge radius and anapole moment models.

We compare the analytic and numerical results for computing the kinematic distributions at LDMX and find that the analytic approach relies on invalid assumptions. The analytic approach assumes that the nucleus is stationary throughout the process, leading to an overestimation in the number of high $|p_{T}|_{e^-_f}$ events.

Gathering enough signal data in the upcoming LDMX and performing model and mass discrimination is an exciting prospect. Having access to both the $p_T$ and $E$ distributions of the final state electron, and in addition the total number of signal events, breaks the degeneracy between the three groups of models we consider in this work: the first with the kinetic mixing, the second with the magnetic and electric dipole, and the third with the charge radius and anapole moment models. We leave a thorough statistical analysis of potential model and mass discrimination of LDMX data to a subsequent study.

\acknowledgments
RC acknowledges support from an individual research grant from the Swedish Research Council (Dnr.~2022-04299). RC and TG have also been funded by the Knut and Alice Wallenberg Foundation, and performed their research within the ``Light Dark Matter'' project (Dnr. KAW 2019.0080).

\appendix
\section{Matrix Elements}
\label{section:matrixelements}
In this appendix we present the matrix elements for each model used in calculating the relic abundance. First we remind the reader of the well known relationship between thermally averaged cross section $\langle \sigma v \rangle_{\chi \bar{\chi} \to f \bar{f}}$ and cross section $\sigma_{\chi \bar{\chi} \to f \bar{f}}$,
\begin{equation}
    \langle \sigma v \rangle_{\chi \bar{\chi} \to f \bar{f}} = \frac{1}{8m_\chi^4 T K_2\left(m_\chi / T \right)^2} \int \sigma(s)_{\chi \bar{\chi} \to f \bar{f}} [s - 4m_\chi^2] \sqrt{s} K_1\left(\sqrt{s}/T \right) ds
\end{equation}
and between the cross section and modulus squared matrix element,
\begin{equation}
    \sigma_{\chi \bar{\chi} \to f \bar{f}} = \frac{|\vec{p}_f|}{64 \pi^2 s |\vec{p}_\chi|} \int \bar{|\mathcal{M}|^2} d\Omega,
\end{equation}
where $T$ is temperature, $K_i$ is the modified Bessel function of the second kind of order $i$, $s$ is the Mandelstam variable, and $|\vec{p}|$ is the magnitude of the 3-momentum.
The squared matrix elements, averaged over initial states and summed over final states, for the relic abundance setting process $\chi \bar{\chi} \to f \bar{f}$, DM annihilation via a virtual $A'$ through an s-channel, are listed here:

\begin{equation}
    \bar{|\mathcal{M}|^2}_{\text{KM}} = \frac{2 g_D^2 \epsilon^2 e^2}{(s - m_{A'}^2)^2 + m_{A'}^2 \Gamma_{A'}^2} (s-4m_{\text{DM}^2})\left[s-(s-4m_f^2)\cos^2\theta \right]
\end{equation}

\begin{equation}
    \bar{|\mathcal{M}|^2}_{\text{M}} = \frac{4 s g_D^2 \mu_f^2}{(s - m_{A'}^2)^2 + m_{A'}^2 \Gamma_{A'}^2} (s-4m_\text{DM}^2) \left[ (s-4m_f^2)\cos^2\theta + 4m_f^2 \right]
\end{equation}

\begin{equation}
    \bar{|\mathcal{M}|^2}_{\text{E}} = \frac{4 s g_D^2 d_f^2}{(s - m_{A'}^2)^2 + m_{A'}^2 \Gamma_{A'}^2} (s-4m_\text{DM}^2) (s-4m_f^2)\cos^2\theta
\end{equation}

\begin{equation}
    \bar{|\mathcal{M}|^2}_{\text{C}} = \frac{4 s^2 g_D^2 b_f^2}{(s - m_{A'}^2)^2 + m_{A'}^2 \Gamma_{A'}^2} (s-4m_\text{DM}^2) \left[ s - (s-4m_f^2)\cos^2\theta \right]
\end{equation}

\begin{equation}
    \bar{|\mathcal{M}|^2}_{\text{A}} = \frac{4 s^2 g_D^2 a_f^2}{(s - m_{A'}^2)^2 + m_{A'}^2 \Gamma_{A'}^2} (s-4m_\text{DM}^2)  (s-4m_f^2)\sin^2\theta
\end{equation}

For the possible decay channels of the dark photon, $A'$, the squared matrix elements are:

\begin{equation}
    \bar{|\mathcal{M}_{A' \to \chi \bar{\chi}}|^2} = \frac{g_D^2}{3} (m_{A'}^2 - 4m_{\text{DM}}^2)
\end{equation}

\begin{equation}
    \bar{|\mathcal{M}_{A' \to f \bar{f}}|^2}_\text{KM} = \frac{e^2 \epsilon^2}{3}(4m_{A'}^2+8m_f^2)
\end{equation}

\begin{equation}
    \bar{|\mathcal{M}_{A' \to f \bar{f}}|^2}_\text{M} = \frac{\mu_f^2}{3}(2m_{A'}^4 + 16m_f^2m_{A'}^2)
\end{equation}

\begin{equation}
    \bar{|\mathcal{M}_{A' \to f \bar{f}}|^2}_\text{E} = \frac{d_f^2}{3}(m_{A'}^4 - 4m_f^2m_{A'}^2)
\end{equation}

\begin{equation}
    \bar{|\mathcal{M}_{A' \to f \bar{f}}|^2}_\text{C} = \frac{b_f^2}{3}(4m_{A'}^6 + 8m_f^2m_{A'}^4)
\end{equation}

\begin{equation}
    \bar{|\mathcal{M}_{A' \to f \bar{f}}|^2}_\text{A} = \frac{a_f^2}{3}(4m_{A'}^6 - 16m_f^2m_{A'}^4)
\end{equation}


\bibliographystyle{JHEP}
\bibliography{biblio.bib}

\end{document}